\documentclass{PoS}

\usepackage{amsmath,amssymb,amsthm}
\usepackage{exscale}
\usepackage{bbm}
\usepackage{tabularx,booktabs,longtable}
\usepackage[T1]{fontenc}
\usepackage{microtype}

\usepackage{graphicx}
\newcommand{\Graph}[2][1.0]{\vcenter{\hbox{\includegraphics[scale=#1]{Graphs/#2}}}}

\usepackage{mathtools}
\usepackage{numprint}
\newcommand{\lfrac}[2]{\tfrac{\numprint{#1}}{\numprint{#2}}}

\usepackage[numbers,sort&compress]{natbib}

\usepackage{xcolor}
\definecolor{links}{rgb}{0,0.3,0}

\usepackage[colorlinks=true,citecolor=blue,urlcolor=links,breaklinks=true,hyperfootnotes=false]{hyperref}

\DeclareMathAlphabet{\oldcal}{OMS}{cmsy}{m}{n}

\newcommand{\field}{\phi}

\newcommand{\lagrangian}{\mathcal{L}}
\newcommand{\MS}{\ensuremath{\mathrm{MS}}}
\newcommand{\bigo}[1]{\oldcal{O}\left(#1\right)}
\newcommand{\D}{D}

\newcommand{\pe}{\nu}

\newcommand{\FI}{\Phi}

\newcommand{\loops}{h_1}

\newcommand{\sdc}{\omega}

\newcommand{\Rbar}{{\R'}\,}
\newcommand{\R}{\oldcal{R}}
\newcommand{\Rstar}{\R^{\ast}}

\newcommand{\Poles}{\K_{\MS}}
\newcommand{\K}{\oldcal{K}}
\newcommand{\onescale}{\text{$1$-s}}
\newcommand{\OneScale}{\K_{\onescale}}

\newcommand{\mzv}[2][]{\zeta^{#1}_{#2}}

\newcommand{\defas}{\mathrel{\mathop:}=}

\newcommand{\set}[1]{\left\{ #1 \right\}}
\newcommand{\abs}[1]{\left\vert#1\right\vert}
\newcommand{\restrict}[2]{\left.{#1}\right|_{#2}}

\newcommand{\U}{\mathcal{U}}
\newcommand{\F}{\mathcal{F}}

\newcommand{\PhiPeriods}[1][]{\ifthenelse{\equal{#1}{}}{\Periods_{\phi^4}}{\Periods_{\phi^4,\leq #1}}}
\newcommand{\LogPeriods}[1][]{\ifthenelse{\equal{#1}{}}{\Periods_{\mathrm{log}}}{\Periods_{\mathrm{log},\leq #1}}}

\newcommand{\Maple}{%
	\href{http://www.maplesoft.com/products/Maple/}
	{\textsf{\textup{Maple}}}%
}
\newcommand{\MapleNote}{%
	\footnote{Maple is a trademark of Waterloo Maple Inc.}
}
\newcommand{\MapleTM}{%
	\href{http://www.maplesoft.com/products/Maple/}
	{\textsf{\textup{Maple}}\texttrademark}%
}
\newcommand{\Python}{%
	\href{https://www.python.org/}
	{\textsf{\textup{Python}}}%
}

\newcommand{\dd}[1][]{\mathrm{d}^{#1}}
\newcommand{\eps}{\varepsilon}

\newcommand{\JaxoDraw}{\texttt{\textup{JaxoDraw}}}
\newcommand{\Axodraw}{\texttt{\textup{Axodraw}}}
\newcommand{\HyperInt}{\texttt{\textup{HyperInt}}}
\newcommand{\GraphState}{\texttt{\textup{GraphState}}}
\newcommand{\Graphine}{\texttt{\textup{Graphine}}}

\title{Renormalization group functions of $\field^4$ theory in the $\MS$-scheme to six loops}

\ShortTitle{Renormalization group functions of $\field^4$ theory in the $\MS$-scheme to six loops}

\author{M.~V.~Kompaniets\\
St.~Petersburg State University\\
7/9 Universitetskaya nab., St.~Petersburg 199034, Russia\\
E-mail: \email{m.kompaniets@spbu.ru}}

\author{\speaker{E.~Panzer}\\
All Souls College, University of Oxford, OX1 4AL Oxford, UK\\
E-mail: \email{erik.panzer@all-souls.ox.ac.uk}}

\abstract{%
	Subdivergences constitute a major obstacle to the evaluation of Feynman integrals and an expression in terms of finite quantities can be a considerable advantage for both analytic and numeric calculations. 
	We report on our implementation of the suggestion by F.~Brown and D.~Kreimer, who proposed to use a modified BPHZ scheme where all counterterms are single-scale integrals.
	Paired with parametric integration via hyperlogarithms, this method is particularly well suited for the computation of renormalization group functions and easily automated.
	As an application of this approach we compute the 6-loop beta function and anomalous dimensions of the $\field^4$ model.
}

\FullConference{%
	Loops and Legs in Quantum Field Theory\\
	24--29 April 2016\\
	Leipzig, Germany%
}

\begin{document}

\section{Introduction}
\label{sec:intro}%

A scalar field with quartic self-interaction is one of the simplest models in quantum field theory. With the discovery of the Higgs boson, it is now part of the standard model of elementary particles. However, it also appears elsewhere and in particular its application as a mean-field approximation to statistical systems is one way to study phase transitions \cite{Vasilev,ZinnJustin,KleinertSchulteFrohlinde:CriticalPhi4}. Critical exponents can be inferred from the renormalization group functions.
Therefore, perturbative calculations in $\field^4$ theory have a long history and they reached a high loop order.

The 5-loop field anomalous dimension and beta function were announced already in \cite{ChetyrkinKataevTkachov:5loopPhi4} and \cite{ChetyrkinGorishnyLarinTkachov:5loopPhi4,Kazakov:MethodOfUniqueness}.
It took ten years until a different group reproduced this calculation and revealed some inaccuracies in \cite{ChetyrkinGorishnyLarinTkachov:5loopPhi4}; corrected results were published in \cite{KNFCL:5loopPhi4}.
A complete numeric check of all these analytic results was completed only much later in \cite{AdzhemyanKompaniets:5loopNumerical}.
Very recently, the first 6-loop result for the field anomalous dimension was published \cite{BatkovichKompanietsChetyrkin:6loop}. 

In this note we present the remaining renormalization group functions of $\field^4$ theory at $6$-loop order, for the minimal subtraction (MS) scheme in dimensional regularization. The results for the $O(N)$-symmetric model and the corresponding critical exponents will be published elsewhere. Instead, we focus here on the technical aspects and the novel tools used in our calculation. 
Note that a completely independent calculation of the $6$-loop beta function with a very different method has recently been finished by Oliver Schnetz \cite{Schnetz:NumbersAndFunctions} and confirmed our result.

We start with the well-known representation of $Z$-factors in terms of massless propagators (also known as \emph{$p$-integrals}). The challenge is then to compute the $\eps$-expansions of such integrals and we refer to \cite{Vasilev,ZinnJustin,KleinertSchulteFrohlinde:CriticalPhi4} for a comprehensive discussion of traditional techniques.
The general philosophy has hitherto been to exploit various relations, in particular the infrared $\Rstar$-operation \cite{ChetyrkinTkachov:InfraredR,ChetyrkinSmirnov:Rcorrected,Chetyrkin:RRR,ChetyrkinGorishnyLarinTkachev:preprint}, to relate the counterterms to simpler integrals. Recently, the cumbersome $\Rstar$-operation was automated in \cite{BatkovichKompaniets:Toolbox}. 
Augmented with the results \cite{BaikovChetyrkin:FourLoopPropagatorsAlgebraic,SmirnovTentyukov:FourLoopPropagatorsNumeric,LeeSmirnov:FourLoopPropagatorsWeightTwelve} for $4$-loop $p$-integrals and integration by parts (IBP) \cite{ChetyrkinTkachov:IBP,VasilevPismakHonkonen:largeNeta}, this traditional approach was used for the $6$-loop calculation of the field anomalous dimension \cite{BatkovichKompanietsChetyrkin:6loop}.
This already required additional tricks for two propagators (see section 5 in \cite{BatkovichKompanietsChetyrkin:6loop}), but the situation is much more severe for the $4$-point diagrams contributing to the $6$-loop beta function: the method fails for $22$ graphs. Of these, $10$ are primitive and have been known for long \citep{BroadhurstKreimer:KnotsNumbers,Schnetz:Census}, but the remaining $12$ graphs contain subdivergences and require new techniques. Initially we calculated these diagrams via parametric integration \cite{Brown:TwoPoint,Panzer:HyperIntAlgorithms} and a new subtraction method for subdivergences (different to the one we elaborate on here), which we will discuss elsewhere.

Instead, in this paper we present a simple new approach based on parametric integration using hyperlogarithms \cite{Brown:TwoPoint,Brown:PeriodsFeynmanIntegrals,Panzer:PhD} which allows us to compute the counterterms of all $6$-loop $2$- and $4$-point graphs analytically, with the sole exception of a single graph which is, however, well-known \cite{Schnetz:K34,Schnetz:GraphicalFunctions,BroadhurstKreimer:KnotsNumbers}.
We use the {\MapleTM} implementation {\HyperInt} \cite{Panzer:HyperIntAlgorithms} of this method, which was presented at the preceding conference \cite{Panzer:LL2014}.\MapleNote
Our new ingredient is an efficient and general procedure to generate convergent integrands for integrals which initially have subdivergences. This is achieved with a BPHZ-like renormalization scheme with one-scale counterterms introduced in \cite{BrownKreimer:AnglesScales}.

A great advantage of this approach is that it is easily automatized and applicable to all integrals, contrary to the multitude of tricks and tools for special classes of diagrams combined in the traditional calculations.
This simplifies the computation, reduces the possibilities for errors in the programs and ensures the reproducibility of our results.%
\footnote{%
	The publication of the {\Maple} and {\Python} programs used for our calculations is under preparation.
}
Also our calculation provides a confirmation of the $5$-loop results \cite{KNFCL:5loopPhi4} and the $6$-loop field anomalous dimension \cite{BatkovichKompanietsChetyrkin:6loop} with very different methods.

\section{Field theory and renormalization}
\label{sec:phi4theory}

The renormalized Lagrangian of the scalar $\field^4$ model in $\D=4-2\eps$ Euclidean dimensions is
\begin{equation}
	\lagrangian
	=
		\frac{1}{2}m^2 Z_1\field^2 
		+\frac{1}{2}Z_2\left(\partial\field \right)^2
		+\frac{16\, \pi^2}{4!}Z_4 \, g\, \mu^{2\eps}\, \field^4
	,
	\label{eq:lagrangian}%
\end{equation}
with an arbitrary mass scale $\mu$. The $Z$-factors relate the renormalized field $\field$, mass $m$ and coupling $g$ to the bare field $\field_0$, bare mass $m_0$ and bare coupling $g_0$ via
\begin{gather}\label{eq:Z-factors}%
	Z_{\field}
		= \frac{\field_0}{\field} 
		= \sqrt{Z_2}
	,\quad
	Z_{m^2} 
		= \frac{m_0^2}{m^2}
		= \frac{Z_1}{Z_2}
	\quad\text{and}\quad
	Z_g
		= \frac{g_0}{\mu^{2\eps} g}
		= \frac{Z_4}{Z_2^2}
	.
\end{gather}
In dimensional regularization \cite{tHooftVeltman:RegularizationGaugeFields} and minimal subtraction, the $Z$-factors depend only on $\eps$ and $g$ as
\begin{equation}
	Z_i = Z_i(g,\eps)
	= 1 + \sum_{k=1}^{\infty} \frac{Z_{i,k}(g)}{\eps^k}
	\label{eq:Z-factor-expansion}%
\end{equation}
and determine the renormalization group functions \cite{tHooft:DimRegRG,Collins:CountertermsDimReg}. We are interested in the beta function
\begin{equation}
	\beta(g)
	= \restrict{
		\mu\frac{\partial g}{\partial \mu}
	}{g_0}
	= - 2\eps \left( \frac{\partial \log (g Z_g)}{\partial g} \right)^{-1}
	= -2\eps g + 2 g^2 \frac{\partial Z_{g,1}(g)}{\partial g}
	\label{eq:beta-def}%
\end{equation}
and the anomalous dimensions for the field and mass, defined by
\begin{equation}
	\gamma_{i}(g) 
	= \restrict{
		\mu \frac{\partial \log Z_{i}}{\partial \mu}
	}{g_0,m_0,\field_0}
	= \beta(g) \frac{\partial \log Z_i(g)}{\partial g}
	= -2 g \frac{\partial Z_{i,1}(g)}{\partial g}
	\quad\text{for}\quad
	i=m^2, \field
	.
	\label{eq:anom-dims}%
\end{equation}
%
%

\section{Counterterms from massless propagators}
\label{sec:counterterms}

The $Z$-factors \eqref{eq:Z-factors} arise as the counterterms for the one-particle irreducible correlation functions $\Gamma_N$ of $N=2$ and $N=4$ fields.
In terms of the Bogoliubov-Parasiuk $\Rbar$-operation \cite{BogoliubovParasiuk:Kausalfunktionen,BogShirk}, which subtracts UV subdivergences from Feynman integrals, the $Z$-factors can be expressed as
\begin{equation}\label{eq:Z-from-R}
\begin{split}
	Z_1 &= 1 + \partial_{m^2} \Poles \Rbar {\Gamma}_2 (p,m^2,g,\mu)
	, \\ 
	Z_2 &= 1 + \partial_{p^2} \Poles \Rbar {\Gamma}_2 (p,m^2,g,\mu)
	\quad\text{and} \\ 
	Z_4 &= 1 + \Poles \Rbar {\Gamma}_4 (p,m^2,g,\mu)/g
	.
\end{split}
\end{equation}
The operator $\Poles$ is the minimal subtraction scheme (\MS), meaning the projection
\begin{equation}
	\Poles \left( \sum_n c_n \eps^n \right)
	\defas
	\sum_{n<0} c_n \eps^n
	\label{eq:Poles-def}%
\end{equation}
onto the pole part with respect to the dimensional regulator $\eps=(4-\D)/2$ and it ensures the form \eqref{eq:Z-factor-expansion} of the $Z$-factors. Recall that these depend only on $g$ and $\eps$, which allows us to simplify the calculation tremendously \cite{Vasilev,ZinnJustin,KleinertSchulteFrohlinde:CriticalPhi4}:
\begin{itemize}
	\item The action of $\partial_{m^2}$ turns a propagator diagram into a sum of diagrams with a squared propagator,
		\begin{equation*}
			\frac{\partial}{\partial m^2} \frac{1}{k^2+m^2}
			= - \frac{1}{k^2+m^2} \frac{1}{k^2+m^2}
		\end{equation*}
		which may be interpreted as a $4$-point graph with two additional, vanishing external momenta entering at a common vertex. This means that $Z_{m^2}$ can be expressed in terms of contributions to $Z_4$ of a subset of the $4$-point diagrams with modified symmetry factors.

	\item We can set all internal masses to zero in the $2$-point diagrams for the computation of $Z_2$, such that $\Gamma_2(p,0,g,\mu)$ is given by $p$-integrals.

	\item Also in $\Gamma_4$ we may set all internal masses to zero. Furthermore, we may set external momenta to zero until only a propagator ($p$-integral) with two external legs remains. This is called infrared rearrangement (IRR) \cite{Vladimirov:ManyLoopPhi4,ChetyrkinKataevTkachov:Gegenbauer}, see figure~\ref{fig:one-scale}.%
\footnote{%
	One may also introduce new (auxiliary) external momenta and it is possible to apply IRR to 2-point diagrams themselves, see the discussion in \cite{BatkovichKompanietsChetyrkin:6loop}. However, we did not use any of these extensions in our calculation.
}
\end{itemize}
These standard techniques express all $Z$-factors in terms of $p$-integrals without non-physical infrared divergences. In the next section, we will explain how these integrals can be computed by parametric integration, at least to the sixth loop order.

As mentioned in the introduction, we will dispose of $\Rstar$ and IBP completely.

\section{Parametric integration}
\label{sec:parametric}

Of the many recent advances made in the evaluation of Feynman integrals, parametric integration is one of the most powerful methods for $p$-integrals; surpassed only by the position-space approach of graphical functions \cite{Schnetz:GraphicalFunctions} and the combination \cite{GolzPanzerSchnetz:GfParam} of both techniques used in \cite{PanzerSchnetz:Phi4Coaction,Schnetz:NumbersAndFunctions}.

The starting point is the representation of a Feynman integral $\FI(G)$ associated to a Feynman graph $G$ in terms of variables $\alpha_e$ associated to each edge $e\in E(G)$ of $G$ (these are called Schwinger-, Feynman- or $\alpha$-parameters).
Let us write $\loops(G)$ for the number of loops of $G$ and
\begin{equation}
	\sdc(G)
	= \sum_{e\in E(G)} \pe_e - \loops(G) \frac{\D}{2}
	= \sum_{e\in E(G)} \pe_e - 2 \loops(G) + \eps \loops(G)
	\label{eq:sdc}%
\end{equation}
for the superficial degree of convergence of $G$ given by power counting. The variables $\pe_e$ encode the exponents to which the momentum space propagators $1/(k_e^2 + m_e^2)^{\pe_e}$ are raised. Choose an arbitrary edge $e_0\in E(G)$.%
\footnote{%
	The value of the Feynman integral $\FI(G)$ in \eqref{eq:UF} does not depend on the choice of $e_0$. It can be interpreted as a projective integral and we can actually replace $\delta(1-\alpha_{e_0})$ by $\delta(1-f(\alpha))$ with an arbitrary function $f(\alpha)$ that is homogeneous of degree one and positive when all $\alpha_e>0$.
}
Then the parametric representation for $\FI(G)$ is
\begin{equation}
	\FI(G)
	= \Gamma(\sdc(G))
	\left(
		\prod_{e\in E(G)}
		\int_0^{\infty}
		\frac{\alpha_e^{\pe_e-1} \dd \alpha_e}{\Gamma(\pe_e)}
	\right)
	\frac{\delta(1-\alpha_{e_0})}{\U^{\D/2-\sdc(G)} \F^{\sdc(G)}}
	\label{eq:alpha-rep}%
\end{equation}
in terms of the Symanzik polynomials $\U$ and $\F$ \cite{Nakanishi:GraphTheoryFeynmanIntegrals,Nakanishi:ParametricAnalyticPerturbation}.
These can be expressed as
\begin{equation}
	\U = \sum_{T} \prod_{e \notin T} \alpha_e
	\quad\text{and}\quad
	\F = \sum_{C} p^2(C) \prod_{e\in C} \alpha_e
	   + \U \sum_{e\in E(G)} m_e^2 \alpha_e
	\label{eq:UF}%
\end{equation}
in terms of spanning trees $T$ and cuts $C$ which separate $G$ into 2 components; $p(C)$ is the momentum flowing through the cut edges $C$ \cite{BognerWeinzierl:GraphPolynomials}.
In our case, all masses $m_e$ are zero and only a single external momentum $p$ is flowing through the graph (it has just $2$ external legs), such that $\F$ (and $\U$) are linear in all edge variables $\alpha_e$. For example, the graph
\begin{equation}
	G = \Graph[0.5]{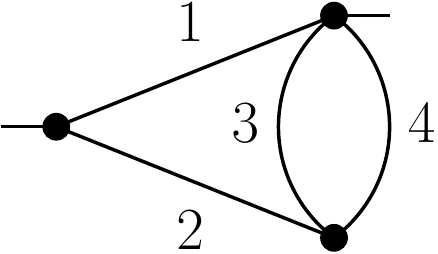}
	\qquad\text{gives}\qquad
	\begin{aligned}
		\U &= (\alpha_1+\alpha_2)(\alpha_3 + \alpha_4) + \alpha_3 \alpha_4
		\quad\text{and}\\
		\F &= p^2 \alpha_1 (\alpha_2 \alpha_3 + \alpha_2 \alpha_4 + \alpha_3 \alpha_4).
	\end{aligned}
	\label{eq:UF-dunce}%
\end{equation}
Note that the momentum dependence factors out of the integral by simple power counting:
\begin{equation}
	\FI(G,p^2) = p^{-2\sdc(G)} \FI(G, 1).
	\label{eq:momentum-dependence}%
\end{equation}
It was noted in \cite{Brown:TwoPoint} that for many massless propagators, the integrals over $\alpha_e$ in \eqref{eq:alpha-rep} can be performed one after the other in terms of multiple polylogarithms if one chooses a good order for these integrations.
Graphs which admit such a good order are called \emph{linearly reducible} and can be computed, order by order in $\eps$, with the algorithm described in \cite{Brown:TwoPoint,Brown:PeriodsFeynmanIntegrals,BognerBrown:GenusZero} and implemented in \cite{Bogner:MPL,Panzer:HyperIntAlgorithms,Panzer:PhD}.
In \cite{Panzer:MasslessPropagators} it was found that all massless propagators with $\leq 4$ loops are indeed linearly reducible, and the same method was even applied to some $6$-loop $p$-integrals \cite{Panzer:LL2014,Panzer:PhD}.

The remaining challenge to the straightforward application of parametric integration in the linearly reducible case is the presence of subdivergences. Note that the $\eps$-expansion needs to be performed on the integrand in \eqref{eq:alpha-rep}. Subdivergences imply that the resulting integrals are divergent and not defined. Therefore we must find integrands which are free of subdivergences.

Sector decomposition \cite{BinothHeinrich:SectorDecomposition,BognerWeinzierl:ResolutionOfSingularities} is the standard approach to solve this problem and by now very well established and powerful \cite{BHJKSZ:SecDec3.0,Smirnov:FIESTA4}. However, it is best suited for numerical calculations; the huge number of sectors that are generated makes it very inefficient for the high loop orders under consideration here, and also it introduces changes of variables which make the analytic integration of each sector much more complicated.

These problems can in principle be avoided with the help of IBP relations, because it is always possible to write a Feynman integral in terms of master integrals without subdivergences \cite{ManteuffelPanzerSchabinger:QuasiFinite}. 
Unfortunately, such IBP reductions are too complicated in our case. Only at this conference, their solution at $4$-loops was achieved \cite{RUV:ForcerLL2016} and highlighted by the impressive computation \cite{BaikovChetyrkinKuehn:5loopQCD} of the $5$-loop QCD $\beta$-function (25 years after the $\field^4$-result). An extension of IBP to the next loop order seems out of reach with current technologies and some new ideas like \cite{ManteuffelSchabinger:Novel} are being investigated.

Luckily, it is possible to avoid IBP altogether by the method explained below. Note that this is feasible only because there are only very few graphs in $\field^4$ theory (just $627$ graphs need to be computed for $\Gamma_4$ at six loops). In contrast, scalar $\field^3$ theory (in six dimensions) has many more diagrams and was therefore evaluated at mere $4$ loops only very recently \cite{Gracey:4loopPhi3,Pismensky:etaphi3in4loop}.

\section{Subdivergences and one-scale renormalization scheme}
\label{sec:one-scale-scheme}

The $\Rbar$-operation subtracts all UV-subdivergences of a given Feynman integral $\FI(G)$ and a final overall subtraction renders the integral itself finite \cite{Hepp:BP}. The famous forest formula \cite{Zimmermann:Bogoliubov}
\begin{equation}
	\R \FI(G)
	= (1-\K) \Rbar \FI(G)
	= \sum_{F} 
		(-1)^{\abs{F}}
		\left[ \FI(G/F) - \K \FI(G/F) \right]
		\prod_{\gamma \in F} \K \FI(\gamma/F)
	\label{eq:forest-formula}%
\end{equation}
expresses the renormalized integral $\R\FI$ explicitly as a sum over all forests $F$ (sets of proper subdivergences which are nested or disjoint).\footnote{%
	Remarkably, this is essentially just a formula for the antipode in the Hopf algebra of Feynman graphs \cite{Kreimer:HopfAlgebraQFT}.
}
The operator $\K$ determines the renormalization scheme and is given by \eqref{eq:Poles-def} for \MS. While \eqref{eq:forest-formula} guarantees that all poles in $\eps$ cancel in the sum of the regularized \emph{integrals}, we do not obtain subdivergence-free \emph{integrands} this way.
The reason is that prior to the integration of the $\alpha$-parameters in \eqref{eq:alpha-rep}, there are no poles in $\eps$ and so we clearly cannot commute the subtraction $\Poles$ with the integration $\FI$. Put differently, by definition of the \MS-scheme, its counterterms depend on the regularization.

This is not the case for schemes like the original BPHZ where the counterterms are defined by expansion, in the masses and momenta, around a fixed set of values (the renormalization point).
It is well-known that in this case one may perform the subtractions under the integral sign; these subtracted integrands give convergent integrals even in $\D=4$ dimensions ($\eps=0$).

Such schemes are therefore ideal for parametric integration and were studied in this context in detail in \cite{BrownKreimer:AnglesScales}.
In particular, the authors suggested a scheme $\OneScale$ where all counterterms are \emph{one-scale}---in other words, $p$-integrals. For logarithmically divergent graphs $G$ they set
\begin{equation}
	\OneScale \FI(G)
	\defas \begin{cases}
		\restrict{\FI(G)}{p^2=1} & \text{if $G$ is a $p$-integral and} \\
		\restrict{\FI(G_{\onescale})}{p^2=1} & \text{if $G$ has more than two external legs,}
	\end{cases}
	\label{eq:onescale-def}%
\end{equation}
\begin{figure}[bt]\ \quad
	$G=\Graph[0.4]{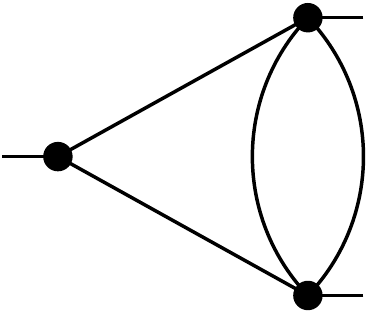}$
	\hfill
	$G_{\onescale}
		= \Graph[0.4]{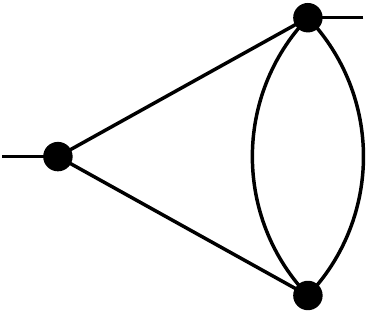}
		\quad\text{or}\quad
		  \Graph[0.4]{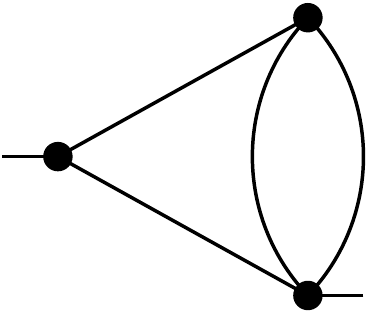}
	$
	\hfill
	$G_{\onescale}^{\text{IR-unsafe}} = \Graph[0.4]{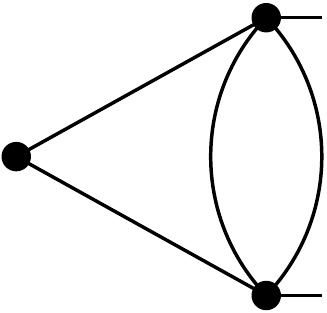}$\quad\ %
	\caption{A graph and its infrared-safe rearrangements. The rearrangement on the right is not infrared-safe.}%
	\label{fig:one-scale}%
\end{figure}%
where $G_{\onescale}$ is any infrared-safe rearrangement of $G$ with only two external legs. We already mentioned this method in section~\ref{sec:counterterms} as the crucial simplification in the computation of $\Rbar \Gamma_4$: There is always at least one way to nullify external momenta to obtain a $p$-integral without introducing infrared divergences, see figure~\ref{fig:one-scale}.
Crucially, the subtractions \eqref{eq:forest-formula} for $\K=\OneScale$ may be performed under the integral sign. We can thus safely expand the subtracted integrand in $\eps$ and integrate each term individually as described in section~\ref{sec:parametric} to obtain the renormalized Feynman integral $\R_{\onescale} \FI(G)$ in this scheme. 
It depends on $p^2$ and vanishes at $p^2=1$ by construction. With \eqref{eq:momentum-dependence}, we find
\begin{equation*}
	\restrict{ 
		\partial_{p^2} \R_{\onescale} \FI(G)
	}{p^2=1}
	= -\sum_{F} (-1)^{\abs{F}} 
		\sdc(G/F) \OneScale \FI(G/F)
		\prod_{\gamma \in F} \OneScale \FI(\gamma/F)
	= - \sdc(G) \OneScale \FI(G) + \ldots
\end{equation*}
and solve for the unrenormalized Feynman integral in dimensional regularization:
\begin{equation}
	\OneScale\FI(G)
	=
	- \frac{\restrict{\left(\partial_{p^2} \R_{\onescale}\FI(G)\right)}{p^2=1}}{\sdc(G)}
	- \sum_{\emptyset \neq F} (-1)^{\abs{F}}
		\frac{\sdc(G/F)}{\sdc(G)} \OneScale\FI(G/F)
		\prod_{\gamma \in G} \OneScale \FI(\gamma/F)
	.
	\label{eq:p-integral-from-one-scale}%
\end{equation}
Note how this formula expresses any given $p$-integral $\FI(G)$ (at $p^2=1$) in terms of
\begin{itemize}
	\item an integral $\R_{\onescale} \FI(G)$ without subdivergences (directly integrable in $\alpha$-parameters) and
	\item products of lower-loop $p$-integrals which can be computed recursively with the same method.
\end{itemize}
An example for the $\Rbar$-operation in this scheme is
\begin{equation}\begin{split}\label{eq:onescale-example}%
	\Rbar_{\onescale} \FI\left( \Graph[0.3]{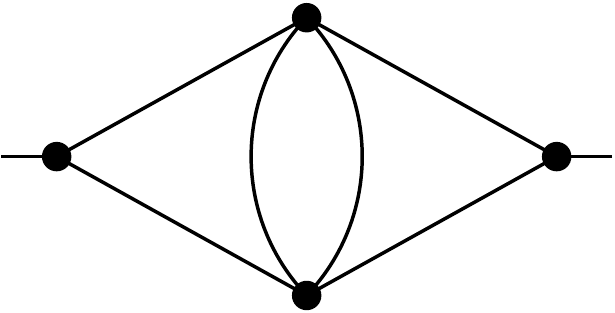} \right)
	&=
	\FI \left( \Graph[0.3]{sauron} \right)
	- \restrict{\FI\left( \Graph[0.4]{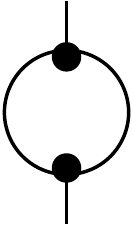} \right)}{p^2=1} 
	\FI\left( \Graph[0.4]{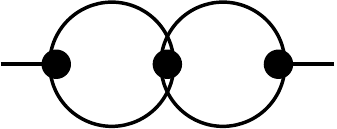}\right)
	\\&\quad
	- \restrict{\FI\left( \Graph[0.3]{dunceleft1} \right)}{p^2=1}
	\FI\left( \Graph[0.4]{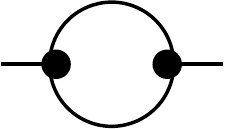}\right)
	- \restrict{\FI\left( \Graph[0.3]{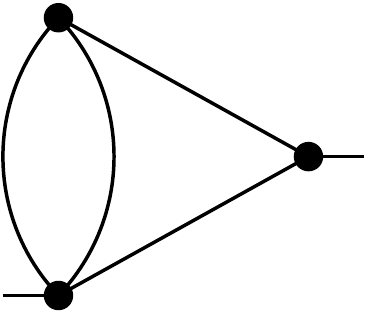} \right)}{p^2=1} 
	\FI\left( \Graph[0.4]{bubblehorz}\right)
	\\&\quad
	+ 2
	\restrict{\FI\left( \Graph[0.35]{bubblevert}\right)}{p^2=1}
	\restrict{\FI\left( \Graph[0.4]{bubblehorz} \right)}{p^2=1} 
	\FI\left( \Graph[0.4]{bubblehorz}\right)
\end{split}\end{equation}
where the terms in the last line come from the forests $F=\set{\Graph[0.2]{bubblevert}, \Graph[0.15]{dunceleft}}$ and $F=\set{\Graph[0.2]{bubblevert}, \Graph[0.15]{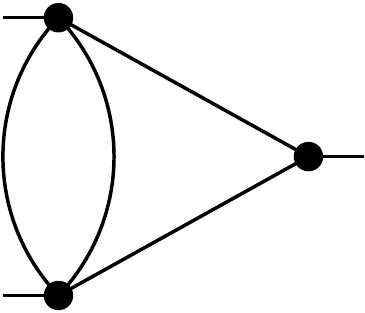}}$. Note that $\partial_{p^2} \R_{\onescale} = \partial_{p^2} \Rbar_{\onescale}$ because $\R_{\onescale} = \Rbar_{\onescale} - \restrict{\Rbar_{\onescale}}{p^2=1}$ differ only by a constant. So we can apply $\partial_{p^2}$ to the left hand side and compute this integral parametrically to the desired order in $\eps$, and then use \eqref{eq:p-integral-from-one-scale} to express the unrenormalized $\FI(\Graph[0.15]{sauron})$ in terms of this quantity and the products of one- and two-loop integrals on the right hand side of \eqref{eq:onescale-example}. In section~\ref{sec:product-polynomials} we discuss a simpler example.

For the overall quadratically divergent 2-point integrals contributing to $\Gamma_2$, the overall subtraction \eqref{eq:onescale-def} becomes $\OneScale\FI(G) = \restrict{\FI(G)}{p^2=1} + (p^2-1) \partial_{p^2}\restrict{\FI(G)}{p^2=1}$. We just take one additional derivative with respect to $p^2$ and compute $\partial_{p^2}^2 \R_{\onescale} = \partial_{p^2}^2 \Rbar_{\onescale}$ in this case, replacing all $\sdc(G)$ in \eqref{eq:p-integral-from-one-scale} by $-\sdc(G) (\sdc(G)+1)$ and similarly for $\sdc(G/F)$.
Note that we can always factor off quadratic \emph{sub}divergences, see section~\ref{sec:factorization}. Hence our above discussion of the simple subtractions for logarithmic subdivergences (together with at most one additional overall derivative) is sufficient for our calculation.
In principle though, the method can also be applied more generally \cite{BrownKreimer:AnglesScales}.

\section{Remarks on the calculation}
\label{sec:calculation}

As was mentioned in the introduction, we wrote computer programs to fully automate the entire calculation. 
The main program is written in {\Python} using the {\GraphState}/{\Graphine} library, which provide a very convenient way to manipulate Feynman graphs \cite{GraphState,BatkovichKompaniets:Toolbox}. It generates the graphs, computes their symmetry factors and combines them into the counterterms \eqref{eq:Z-from-R} to calculate the anomalous dimensions and the beta function via \eqref{eq:anom-dims} and \eqref{eq:beta-def}. 
In order to evaluate the corresponding integrals $\FI(G)$, the {\Python} program computes the forest formula \eqref{eq:forest-formula} and chooses one-scale structures for each graph, implementing the scheme \eqref{eq:onescale-def}.

The resulting expressions of $\R_{\onescale} \FI(G)$ as linear combinations of products of integrals are passed on to a {\Maple} script. It computes the parametric representation according to section~\ref{sec:product-polynomials}, performs the $\eps$-expansion and calls {\HyperInt} \cite{Panzer:HyperIntAlgorithms} to perform the integration of $\restrict{\left(\partial_{p^2} \R_{\onescale}\FI(G)\right)}{p^2=1}$. This is the most time-consuming step.
Finally, the {\Python} program reconstructs the value of the unsubtracted $p$-integral according to \eqref{eq:p-integral-from-one-scale}. 

A detailed discussion and publication of these programs is under preparation.

\subsection{Factorization}
\label{sec:factorization}

Due to the trivial momentum dependence \eqref{eq:momentum-dependence} of $p$-integrals, every $2$-point subgraph $\gamma$ can be integrated out separately if one replaces it with a propagator of index $\pe_e=\sdc(\gamma)$. For example,
\begin{equation*}
	\FI\left( \Graph[0.8]{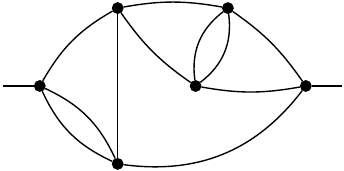} \right)
	= \restrict{\FI\left( \Graph{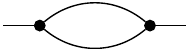} \right)^2}{p^2=1}
	\cdot
	\restrict{ \FI \left(\Graph{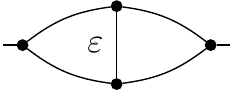} \right)}{p^2=1}
	\cdot
	\FI \left( \Graph{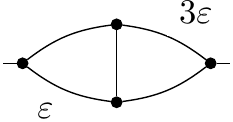} \right)
\end{equation*}
Therefore, we only need to compute $p$-integrals for $2$-connected graphs (graphs which do not have any $p$-integral as a subgraph). In particular this means that all quadratic propagator subdivergences factor out and we only have to deal with logarithmic subdivergences, where the simple subtractions \eqref{eq:onescale-def} suffice.

Note that this factorization procedure introduces $\eps$-dependent propagator exponents $\pe_e$. With traditional methods it was very difficult to compute such integrals. For parametric integration, the effect of such exponents is just that the $\eps$-expansion of the integrand in \eqref{eq:alpha-rep} also includes logarithms $\log (\alpha_e)$ of individual Schwinger parameters, in addition to $\log(\U)$ and $\log(\F)$. This causes no problem for the integration algorithms. For example, comprehensive results for \emph{arbitrary} $\pe_e$ in the case of $4$-loop $p$-integrals were computed this way in \cite{Panzer:MasslessPropagators}.

\subsection{Parametric representation for products}
\label{sec:product-polynomials}

The forest formula \eqref{eq:forest-formula} for $\R_{\onescale} \FI(G)$ contains products of Feynman integrals. Note that the graph polynomials as defined in \eqref{eq:UF} vanish for products (disjoint unions) of graphs. Instead, for the parametric integral representation for a product $G = \prod_{i=1}^n G_i$ one has to set \cite{BrownKreimer:AnglesScales}
\begin{equation}
	\U_G = \prod_{i=1}^n \U_{G_i}
	\quad\text{and}\quad
	\F_G = \sum_{i=1}^n \F_{G_i} \prod_{j\neq i}\U_{G_j}
	\quad\text{such that}\quad
	\frac{\F_G}{\U_G}
	= \sum_{i=1}^n \frac{\F_{G_i}}{\U_{G_i}}
	.
	\label{eq:UF-product}%
\end{equation}
In order to ensure the cancellation of all subdivergences in the parametric integral representation for $\R_{\onescale} \FI(G)$, it is crucial that one tracks correctly the individual edges of the sub- and quotient graphs in \eqref{eq:forest-formula}. This is illustrated in the following example of a single logarithmic UV-subdivergence from \eqref{eq:UF-dunce}, where we explicitly show the edge labels:
\begin{equation}
	\Rbar_{\onescale} \FI \left( \Graph[0.45]{dunce_1s} \right)
	= \FI\left( \Graph[0.45]{dunce_1s} \right)
	- \restrict{\FI\left( \Graph[0.45]{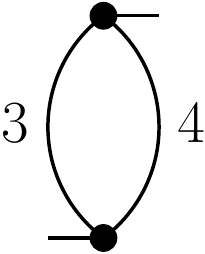} \right)}{p^2=1}
	\FI\left( \Graph[0.45]{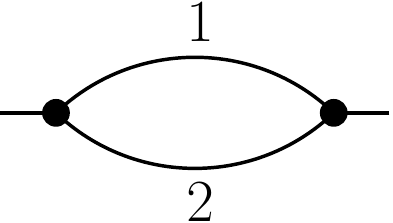} \right)
	.
	\label{eq:dunce-labels}%
\end{equation}
The graph polynomials \eqref{eq:UF} for the graph $G=\Graph[0.2]{dunce_1s}$ were given in \eqref{eq:UF-dunce} and the prescription \eqref{eq:UF-product} for the product of the subgraph $\gamma=\Graph[0.2]{dunce_1s_sub}$ at $p^2=1$ and the quotient $G/\gamma=\Graph[0.2]{dunce_1s_co}$ yields
\begin{equation}
	\U_{\gamma\cdot G/\gamma}
	= (\alpha_1+\alpha_2)(\alpha_3+\alpha_4)
	\quad\text{and}\quad
	\F_{\gamma \cdot G/\gamma}
	= \alpha_3 \alpha_4 (\alpha_1+\alpha_2)
	+ p^2 \alpha_1 \alpha_2 (\alpha_3 + \alpha_4)
	.
	\label{eq:UF-dunce-product}%
\end{equation}
The parametric representation \eqref{eq:alpha-rep} for $\restrict{\partial_{p^2} \R_{\onescale} \FI(G)}{p^2=1}$ in this case is
\begin{equation}
	\Gamma(2\eps) \int_0^\infty \dd \alpha_1 \cdots \int_0^{\infty} \dd \alpha_4
	\delta(1-\alpha_{e_0})
	\left[
		\frac{-2\eps}{\U_G^{2-3\eps} \F_G^{2\eps}}
		- \frac{-\eps \alpha_1 \alpha_2(\alpha_3 + \alpha_4)}{\U_{\gamma \cdot G/\gamma}^{2-2\eps} \F_{\gamma \cdot G/\gamma}^{1+\eps}}
	\right]_{p^2=1}
	\label{eq:dunce-parametric}%
\end{equation}
which is finite at $\eps\rightarrow 0$ and may be integrated order by order in $\eps$.
Many more examples are discussed in \cite{BrownKreimer:AnglesScales}.

\subsection{Simplification at leading order}

For the leading order in $\eps$, we evaluate the parametric integrand in \eqref{eq:alpha-rep} at $\eps=0$.
If a graph $G$ is logarithmically divergent, that is $\restrict{\sdc(G)}{\eps=0}=0$, this means that $\F$ drops out completely and only the polynomial $\U$ plays a role for the integration. It determines the pole in $\eps$ coming from the $\Gamma(\sdc)$-prefactor; higher orders in $\eps$ are not needed for the determination of the $Z$-factors.
This simplifies the parametric integration considerably. Since $\U$ is linear in all $\alpha_e$, the first integration is elementary and the result can be interpreted as the $p$-integral $G\setminus e$ (with the external momenta entering at the vertices that were incident to $e$) \cite{Panzer:MasslessPropagators,Brown:TwoPoint}.\footnote{%
	These relations between different $p$-integrals and vacuum integrals are also known as \emph{cut-and-glue} \cite{BaikovChetyrkin:FourLoopPropagatorsAlgebraic}.
}

All $4$-point $6$-loop graphs contributing to $\Gamma_4$ are therefore effectively expressed in terms of $5$-loop $p$-integrals. In the traditional approach, this simplification is achieved with the $\Rstar$-operation \cite{ChetyrkinTkachov:InfraredR,ChetyrkinSmirnov:Rcorrected,Chetyrkin:RRR,ChetyrkinGorishnyLarinTkachev:preprint}.

\subsection{Linear reducibility}

Not all $6$-loop $p$-integrals are linearly reducible, but as explained we are essentially only computing $5$-loop $p$-integrals. It turns out that indeed all integrals needed for the $6$-loop calculation of the $Z$-factors are linearly reducible, with just a single exception.
This is the primitive graph
\begin{equation}
	\FI\left( \Graph[0.2]{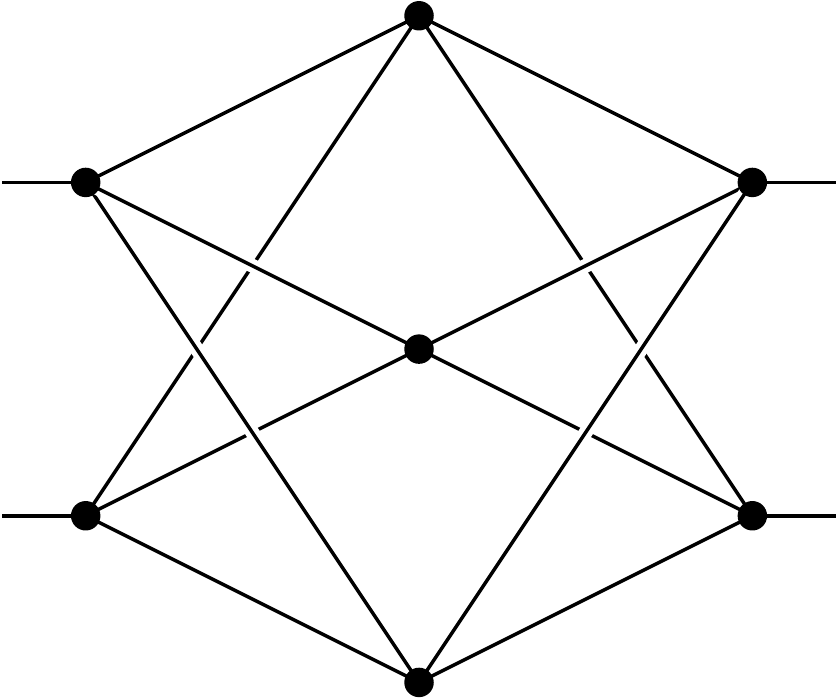} \right)
	= \frac{288}{30\eps} \Big(
		58 \mzv{8} - 45 \mzv{3} \mzv{5} - 24 \mzv{3,5}
	\Big)
	+ \bigo{\eps^0},
	\label{eq:P64}%
\end{equation}
$P_{6,4}$ in the notation of \cite{Schnetz:Census}. It has been known numerically since \cite{Broadhurst:5loopsbeyond} and in \cite{BroadhurstKreimer:KnotsNumbers} it was identified in terms of Riemann zeta values $\mzv{k} = \sum_{n=1}^{\infty} 1/n^k$ and the double zeta value
\begin{equation}
	\mzv{3,5}
	\defas
	\sum_{1\leq n < m} \frac{1}{n^3 m^5}
	\approx 
	0.037707673
	.
	\label{eq:zeta35}%
\end{equation}
It is conjectured that $\mzv{3,5}$ cannot be expressed as a rational linear combination of products of Riemann zeta values.
An analytic calculation confirming the result \eqref{eq:P64} was first provided in \cite{Schnetz:K34} and recently with the beautifully elegant method of graphical functions \cite{Schnetz:GraphicalFunctions}.
While $P_{6,4}$ is not linearly reducible in the strict sense and thus not computable with {\HyperInt}, this is in fact only a limitation of this implementation which could be lifted: It is known how $P_{6,4}$ can be integrated parametrically by splitting the integrand as described in \cite{Brown:TwoPoint}.

\section{Results and checks}
\label{sec:result}

The beta function of $\field^4$-theory in the $\MS$ scheme in $\D=4-2\eps$ dimensions to order $g^7$ is
\begin{equation}\begin{split}
\label{eq:beta-result}%
	\beta^{\MS}(g) &=
-2 \varepsilon g
+3 g^2
- \lfrac{17}{3} g^3
+\left(
    12 \mzv{3}
    +\lfrac{145}{8} 
\right) g^4
-\left(
    120 \mzv{5}
    -18 \mzv{4}
    +78 \mzv{3}
    +\lfrac{3499}{48} 
\right) g^5
\\&\quad
+\left(
    \numprint{1323} \mzv{7}
    +45 \mzv[2]{3}
    -\lfrac{675}{2} \mzv{6}
    +987 \mzv{5}
    -\lfrac{1189}{8} \mzv{4}
    +\lfrac{7965}{16} \mzv{3}
    +\lfrac{764621}{2304} 
\right) g^6
\\&\quad
-\left(
    \lfrac{46112}{3} \mzv{9}
    +768 \mzv[3]{3}
    +\lfrac{51984}{25} \mzv{3,5}
    -\lfrac{264543}{25} \mzv{8}
    +\numprint{4704} \mzv{3}\mzv{5}
    +\lfrac{63627}{5} \mzv{7}
    -162 \mzv{3}\mzv{4}
\right.\\&\left.\qquad\ 
    +\lfrac{8678}{5} \mzv[2]{3}
    -\lfrac{6691}{2} \mzv{6}
    +\lfrac{63723}{10} \mzv{5}
    -\lfrac{16989}{16} \mzv{4}
    +\lfrac{779603}{240} \mzv{3}
    +\lfrac{18841427}{11520} 
\right) g^7
+\bigo{g^8} 
\\&\approx
-2 \epsilon g
+3 g^2
-5.667 g^3
+32.55 g^4
-271.6 g^5
+2849 g^6
-34776 g^7
+\bigo{g^8}
.
\end{split}\end{equation}
For the anomalous dimension of the mass to order $g^6$ we find
\begin{align}
	\gamma^{\MS}_{m^2}(g) &=
-g
+\lfrac{5}{6} g^2
-\lfrac{7}{2} g^3
+\left(
    3 \mzv{4}
    +\lfrac{3}{2} \mzv{3}
    +\lfrac{477}{32} 
\right) g^4
-\left(
    \lfrac{75}{2} \mzv{6}
    -9 \mzv[2]{3}
    + \mzv{5}
    +\lfrac{65}{4} \mzv{4}
    +\lfrac{1519}{48} \mzv{3}
    +\lfrac{158849}{2304} 
\right) g^5
\nonumber\\&\quad
+\left(
    \lfrac{55701}{100} \mzv{8}
    -288 \mzv{3}\mzv{5}
    -\lfrac{972}{25} \mzv{3,5}
    +54 \mzv{3}\mzv{4}
    -\lfrac{4629}{20} \mzv{7}
\right.\nonumber\\&\left.\qquad\ 
    +\lfrac{446}{5} \mzv[2]{3}
    +\lfrac{1141}{4} \mzv{6}
    +\lfrac{4019}{40} \mzv{5}
    +\lfrac{1695}{32} \mzv{4}
    +\lfrac{472891}{1440} \mzv{3}
    +\lfrac{7915913}{23040} 
\right) g^6
+\bigo{g^7} 
\label{eq:gamma-m2-result}%
\\& \approx 
-g
+0.8333 g^2
-3.5 g^3
+19.96 g^4
-150.8 g^5
+1355 g^6
+\bigo{g^7}
.\nonumber
\end{align}
Let us summarize various checks of our result.
First note that we computed the renormalization group functions for the $O(N)$-symmetric $\field^4$ model with their full $N$-dependence (shown above for $N=1$). Our result agrees with the known $5$-loop results \cite{ChetyrkinKataevTkachov:5loopPhi4,ChetyrkinGorishnyLarinTkachov:5loopPhi4,Kazakov:MethodOfUniqueness,KNFCL:5loopPhi4} and the 6-loop field anomalous dimension \cite{BatkovichKompanietsChetyrkin:6loop}.
It also confirms the leading and subleading terms in the large $N$ expansion of the $6$-loop beta function obtained almost 20 years ago by John Gracey in \cite{Gracey:LargeNf}.

Additional checks were provided by Dmitrii Batkovich, who calculated $575$ out of the $627$ six-loop $\Gamma_4$-counterterms using the traditional approach, which combines IBP for four-loop massless propagators (known from \cite{BaikovChetyrkin:FourLoopPropagatorsAlgebraic,SmirnovTentyukov:FourLoopPropagatorsNumeric,LeeSmirnov:FourLoopPropagatorsWeightTwelve}) and IRR with $\Rstar$ to compute the diagrams \cite{BatkovichKompaniets:6loop-Rstar}.
The most complicated diagrams (that could not be checked this way) were additionally computed numerically with sector decomposition \cite{BinothHeinrich:SectorDecomposition}, using a custom-made program by the first author, to at least 3 significant digits.

Also note that in our first calculation, we resolved subdivergences in a different way than presented in section~\ref{sec:one-scale-scheme}, namely by constructing primitive (i.e.\ subdivergence-free) linear combinations of graphs like in \cite{Panzer:MasslessPropagators,Panzer:PhD}. This technique will be explained in detail elsewhere. The results obtained by both methods agree.

Probably the strongest check comes from the completely independent computation of the $6$-loop renormalization group functions of $\field^4$ theory ($N=1$) by Oliver Schnetz \cite{Schnetz:NumbersAndFunctions}. His computation is very different both conceptually and technically; being carried out in position space with graphical functions \cite{Schnetz:GraphicalFunctions} and exploiting the powerful theory of generalized single-valued hyperlogarithms \cite{Schnetz:GeneralizedSV}.

\section{Outlook}
\label{sec:outlook}

While the primitive $6$-loop graphs had been computed already thirty years ago \cite{Broadhurst:5loopsbeyond}, subdivergences have hitherto been the obstacle for calculations in $\field^4$ theory. In this article, we showed how recently developed techniques (parametric integration and the one-scale BPHZ scheme) can overcome the limitations of traditional methods.
This approach can principle also be used at $7$ loops, but the calculation using graphical functions promises to be much more efficient and is already underway \cite{Schnetz:NumbersAndFunctions}.
The contributions from $7$-loop graphs without subdivergences (these are expected to give the most complicated transcendental numbers) have been completed already \cite{PanzerSchnetz:Phi4Coaction}.

An obvious question is to which extent the techniques presented here are will be of use in other theories, like $\field^3$ in six dimensions (only known to four loops \cite{Gracey:4loopPhi3}) and gauge theories like QCD (known to five loops \cite{BaikovChetyrkinKuehn:5loopQCD}). One problem is the much higher number of graphs at a given loop order.
A further complication might arise from the integrals themselves, because cubic graphs have more edges per loop order than graphs of $\field^4$. This means that the integrations will be more demanding.

We also hope that graph-theoretic methods of desingularization, for example the BPHZ-derived approach we described here, might be useful in more general situations when the method \cite{ManteuffelPanzerSchabinger:QuasiFinite} via IBP is not an option.

\acknowledgments
Both authors wish to express their admiration and sincere thanks towards Kostja Chetyrkin. Apart from his numerous impressive achievements, he is a very helpful and open researcher who shares his knowledge, expertise and experience. He kindly invited the second author to \href{https://www.ttp.kit.edu/}{KIT} in 2014 where he introduced him to the first author and encouraged us to undertake the $6$-loop calculation.

We are deeply indebted to Oliver Schnetz for providing us with the result of his own, independent computation in February of this year. At that time, the method presented here was not yet implemented and hence we had computed the majority of diagrams only via the traditional IBP/IRR/$\Rstar$-method (back then, we used parametric integration only for the most complicated diagrams). We were thus anxious to find a second method to evaluate the diagrams to exclude the possibility of errors.
The comparison with Oliver's result convinced us of the correctness, since the only discrepancy was localized in the coefficient of $\mzv{9}$ and could be tracked down to a copy error when we transferred the results for the primitive graphs from \cite{Schnetz:Census} into our program.%
\footnote{
	Since then we implemented the new method described in this article and recomputed all integrals this way, finding perfect agreement. This highly automated program computes all integrals from scratch, with the sole exception of \eqref{eq:P64}.
}

Very valuable checks of plenty of $6$-loop integrals via the IBP/IRR/$\Rstar$-method where provided by Dmitrii Batkovich \cite{BatkovichKompaniets:6loop-Rstar}.

Also we thank John Gracey for providing us with his results \cite{Gracey:LargeNf} on the large $N$ expansion. This check was performed during this conference, and we like to thank the organizers of Loops~\&~Legs 2016 for the chance to present our result and bringing together so many exceptional scientists in this forum. The participants made it a very stimulating event, fostering scientific exchange.

David Broadhurst provided us with a copy of \cite{Broadhurst:WithoutSubtractions}, in which he computes the $\overline{\MS}$-renormalized $5$-loop propagator of $O(N)$-symmetric $\field^4$ theory with ingenuity and extreme economy. It provided additional strong checks of our result, since it amounts to an evaluation of the finite parts of $5$-loop propagator diagrams, which in turn give a subset of the counterterms at $6$ loops.

Most figures in this article were created with {\JaxoDraw} \cite{BinosiTheussl:JaxoDraw} and {\Axodraw} \cite{Vermaseren:Axodraw}.

\bibliographystyle{JHEPsortdoi}
\bibliography{refs}

\end{document}